\documentclass[conference]{IEEEtran}
\usepackage{orcidlink}
\IEEEoverridecommandlockouts
\usepackage{cite}
\usepackage{amsmath,amssymb,amsfonts}
\usepackage{algorithmic}
\usepackage{graphicx}
\usepackage{textcomp}
\usepackage{xcolor}
\usepackage{tcolorbox}
\def\BibTeX{{\rm B\kern-.05em{\sc i\kern-.025em b}\kern-.08em
    T\kern-.1667em\lower.7ex\hbox{E}\kern-.125emX}}
\usepackage{listings}
\usepackage{wrapfig}
\usepackage{subcaption}
\usepackage{bbding}
\usepackage{stfloats} 
\usepackage{tikz}
\usepackage{url}
\usepackage{minted}
\usepackage{enumitem}

\usepackage{booktabs}
\usepackage{multirow}

\usepackage{pifont}

\begin{document}

\title{
The Phish, The Spam, and The Valid: Generating Feature-Rich Emails for Benchmarking LLMs
\\}

\author{
    \IEEEauthorblockN{Rebeka Toth}
    \IEEEauthorblockA{\textit{Department of Informatics}\\
    \textit{University of Oslo} \\
    Oslo, Norway \\
    \orcidlink{0009-0000-9574-1896}}
    \and
    \IEEEauthorblockN{Nils Gruschka}
    \IEEEauthorblockA{\textit{Department of Informatics}\\
    \textit{University of Oslo} \\
    Oslo, Norway \\
    \orcidlink{0000-0001-7360-8314}}
    \and
    \IEEEauthorblockN{Tamas Bisztray$^*$}
    \IEEEauthorblockA{\textit{Department of Informatics}\\
    \textit{University of Oslo} \\
    Oslo, Norway \\
    \orcidlink{0000-0003-2626-3434}
    \thanks{$^*$Corresponding author: tamasbi@uio.no}} 
}

\maketitle

\begin{abstract}

In this paper, we introduce a metadata-enriched generation framework (PhishFuzzer) that seeds real emails into Large Language Models (LLMs) to produce 23,100 diverse, structurally consistent email variants across controlled entity and length dimensions. Unlike prior corpora, our dataset features strict three-class labels (Phishing, Spam, Valid), provides full URL and attachment metadata, and annotates each email with attacker intent. Using this dataset, we benchmark two state-of-the-art LLMs (Qwen-2.5-72B and Gemini-3.1-Pro) under both Basic (body, subject) and Full (+URL, sender, attachment) settings. By applying formal confidence metrics (Task Success Rate and Confidence Index), we analyze model reliability, robustness against linguistic fuzzing, and the impact of structural metadata on detection accuracy. Our fully open-source framework and dataset provide a rigorous foundation for evaluating next-generation email security systems. To support open science, we make the PhishFuzzer Dataset, the generation scripts and prompts available on GitHub: https://github.com/DataPhish/PhishFuzzer
\end{abstract}

\begin{IEEEkeywords}
phishing, spam, email security, large language models, email detection, emotion analysis, dataset
\end{IEEEkeywords}

\section{Introduction}
Phishing and spam emails continue to pose a significant threat to cybersecurity, targeting individuals and organizations with deceptive messages designed to steal sensitive information, compromise systems, and facilitate fraudulent activity~\cite{enisa2025}. According to the ENISA Threat Landscape 2025 report, phishing remains one of the dominant initial access vectors across major incident categories, with continued growth in both volume and sophistication~\cite{enisa2025}.

Cybercriminals increasingly exploit Large Language Models (LLMs) to produce highly convincing and linguistically polished messages at scale. Recent work shows that LLM-generated phishing can achieve click-through rates comparable to human-crafted attacks~\cite{10466545}, while traditional detection systems that utilize rule-based heuristics and static features such as keywords, sender reputation, or metadata, degrade significantly when confronted with LLM-rephrased content~\cite{afane2024nextgenerationphishingllmagents,10.1145/3733799.3762967}. This highlights the need for more robust and adaptive email security solutions. 

Meanwhile, existing datasets are often outdated, lack separate classes for phishing, spam and valid emails, metadata and linguistic variance required for modern email classification.

\noindent
To bridge this critical gap in dataset quality and model evaluation, we make the following contributions:

\begin{itemize}
    \item \textbf{PhishFuzzer}: An LLM-based generation pipeline that uses real-world emails as templates to produce structurally consistent synthetic variants.
    \item \textbf{PhishFuzzer Dataset}: The first open-source, three-class (Phishing, Spam, Valid) email dataset comprising 3,300 real seeds and 19,800 synthetic variants, complete with attacker motivation annotations, structural metadata, and strict provenance tracking.
    \item \textbf{Rigorous LLM Benchmarking:} A comprehensive evaluation of Qwen-2.5-72B and Gemini-3.1-Pro demonstrating that while LLMs achieve high zero-shot phishing detection, their performance is sensitive to the inclusion of metadata and they struggle with the subjective boundary between spam and valid email.
    \item \textbf{Systematic Failure Analysis:} Through the introduction of the Total Flip Score (TFS@K) metric, we isolate model blind spots, and inaccuracies in human labels in legacy public datasets.
\end{itemize}

This paper is structured as follows: Section~\ref{sec:related} discusses related work, Section~\ref{sec:methodology} presents our methodology, Section~\ref{sec:results} presents our results, Section~\ref{sec:limit} discusses limitations and future work, while Section~\ref{sec:conclusion} summarizes and concludes our work.

\section{Related Literature}
\label{sec:related}

\subsection{Existing Email Datasets}
Table~\ref{tab:dataset_comparison} summarizes the properties of existing open-source email corpora, highlighting several critical shortcomings. Most widely used datasets were collected before 2010 and frequently suffer from data quality issues such as encoding inconsistencies, malformed characters, and residual HTML artifacts. Because of their age, they do not reflect the improved linguistic quality and structural sophistication seen in modern AI-assisted campaigns~\cite{10.1145/3733799.3762967,10466545}. Furthermore, some of these legacy datasets strip away important real-world signals---including full URL structures, attachment names, and sender-domain features---and offer no annotations regarding the attacker's underlying intent. 
\begin{table}[t]
\centering
\caption{Comparison of phishing datasets based on granularity, origin, and availability.}
\label{tab:dataset_comparison}
\begin{tabular}{@{}l@{\hspace{4pt}}r@{\hspace{4pt}}c@{\hspace{4pt}}c@{\hspace{6pt}}c@{\hspace{6pt}}c@{\hspace{6pt}}c@{\hspace{6pt}}c@{}}
\toprule
\textbf{Dataset} & \textbf{Size} & \textbf{Year} & \textbf{Classes} & \textbf{Metad.} & \textbf{Intent} & \textbf{Real/Syn.} & \textbf{Open} \\
\midrule
SpamAssas.~\cite{spamassassin_corpus} & 6k     & 2003 & S\&V$^*$ & N & N & Real & Y \\
Enron~\cite{Kaggle}           & 517k   & 2004 & -- & Y & N & Real & Y \\
Nazario~\cite{Kaggle}         & 11k    & $\leq$21 & P & N & N & Real & Y \\
CEAS-08~\cite{Kaggle}         & 39k    & 2008 & S\&V & N & N & Real & Y \\
Nig.Fraud~\cite{Kaggle}       & 4k     & $\leq$07 & P & N & N & Real & Y \\
Codebook~\cite{saka2024phishing}      & 503    & $\leq$21 & P & N & Y\textsuperscript{**} & Real & N \\
E-PhishLLM~\cite{10.1145/3733799.3762967}  & 16k    & 2025 & P\&V & N & N & Syn & Y \\
\midrule
\textbf{PhishFuzzer}         & 23k & $\leq$26 & P\&S\&V & Y & Y & Both & Y \\
\bottomrule
\end{tabular}

\vspace{2pt}
\noindent V = valid emails, S = spam, P = phishing \\
$^*$Multiple difficulty levels are defined, but they are arbitrary.\\
$^{**}$Unclear if dataset is annotated, not released.
\end{table}


To address the lack of modern data, Pajola et al.\ proposed E-PhishGEN, a framework for generating fully synthetic phishing and valid emails from LLM-created user profiles~\cite{10.1145/3733799.3762967}. Their cross-dataset evaluation revealed that classical machine learning (ML) approaches trained on legacy corpora and tested on E-PhishLLM experienced accuracy drops of up to 40 percentage points, signaling that legacy datasets are insufficient proxies for LLM generated phishing content.
However, their cross-training experiments were limited to classical ML pipelines, omitting more advanced transformer-based models. Furthermore, E-PhishLLM lacks the structural granularity such as explicit URL strings, attachment names, that can be critical for both human and algorithmic decision-making.

\subsection{LLM-Based Email Classification}

Recent research has started examining phishing intent. Eilertsen et al.~\cite{eilertsen2025llm} proposed an intent-based phishing taxonomy derived from MITRE ATT\&CK T1566, categorizing emails as \textit{Phishing via Link}, \textit{Attachment}, \textit{Service}, or \textit{Other}. While highly valuable for threat modeling, this categorization relies entirely on manual annotation.

Saka et al.~\cite{saka2024phishing} manually categorized 503 emails from the Nazario dataset based on the specific actions requested from the victim---‘click’, ‘download’, ‘reply/email’, ‘call’, ‘other’, and ‘none’---however, they also did not provide an automated or LLM-based method to extract these labels at scale.

A growing amount of work has benchmarked LLMs for phishing and spam detection, with GPT-4o, Gemini 1.5, Llama-3.1, and Mistral-Large matching or exceeding fine-tuned transformer baselines~\cite{10466545,10878987,Mardiansyah2024,Mahendru2024,Lee2025}. In parallel, studies on LLM-generated phishing show that GPT-4-crafted emails rival human-written attacks in persuasiveness~\cite{10466545}. Afane et al.~\cite{afane2024nextgenerationphishingllmagents} found that detection accuracy drops substantially under LLM-based rephrasing when tested against traditional phishing detectors (e.g., Gmail Spam Filter, Apache SpamAssassin, Proofpoint) as well as classical machine learning models like SVM and Logistic Regression. These findings underscore a critical shift: as traditional filters fail against LLM-rephrased attacks, robust detection will increasingly rely on advanced architectures, ranging from fine-tuned local transformers (e.g., BERT) to zero-shot reasoning models (e.g., GPT-4o). To effectively train and benchmark these systems, the community requires realistic, large-scale datasets that preserve structural metadata and attacker intent. Our generation methodology directly addresses this need by resolving the fundamental shortcomings of existing approaches highlighted in Table~\ref{tab:dataset_comparison}.

\section{Methodology}
\label{sec:methodology}

\begin{figure*}[ht]
    \centering
    \includegraphics[width=0.94\linewidth]{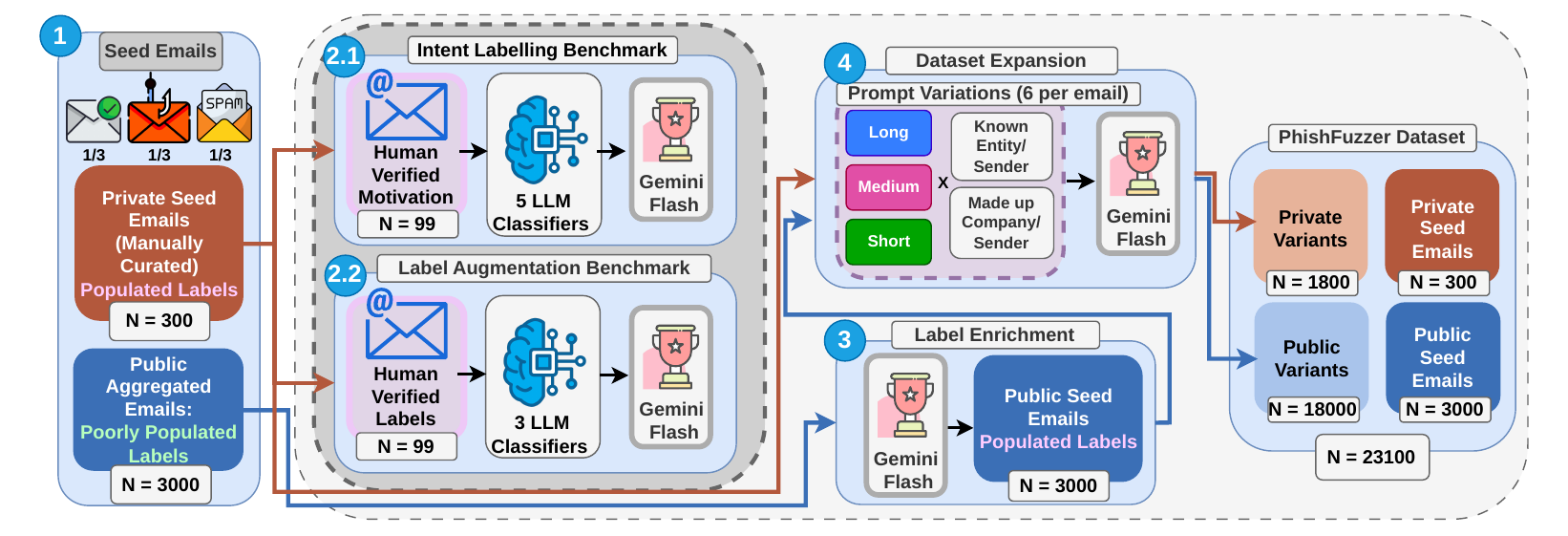}
    \caption{Methodology for the PhishFuzzer Framework}
    \label{fig:methodology}
\end{figure*}

This section provides an overview of both the dataset creation pipeline, and the LLM benchmarking approach.
Figure~\ref{fig:methodology} provides an overview of the four steps of the dataset creation. 
\subsection{Ethical Guidelines} 
This research was conducted in strict accordance with the ethical guidelines of the University of Oslo and Norwegian national regulations. All participants provided informed consent, and the submitted data underwent strict manual de-identification to ensure full anonymity.

\subsection{Step 1: Seed Dataset Creation}
\label{sec:dataset_creation}
The seed dataset combines two complementary sources:
\textbf{Manually Curated Private (N = 300):} 
Constructed from anonymized emails voluntarily submitted from personal and corporate inboxes, balanced across phishing (including enterprise phishing awareness campaigns~\cite{toth2025}), spam, and legitimate (100 each). Each was annotated with subject, body, sender, all extracted URLs, and attachment filenames with extensions. 
\noindent
\textbf{Aggregated Public (N = 3,000):} Additional emails were drawn from public datasets~\cite{Kaggle} and \cite{spamassassin_corpus}. These follow a coarser schema: URLs appear only when present as plaintext in the body, and attachments are recorded as binary flags rather than by filename. Steps~2--3 address this reduced granularity.

\subsection{Step 2.1: Intent Benchmarking}
\label{sec:motivation_benchmarking}

We benchmarked LLMs on identifying the \textit{primary explicitly requested user action} in an email, not equivalent to the mere presence of an artifact.

A subset of 99 emails (33 per class) was independently labeled by two domain experts with one of four intent categories: \textit{Follow the link}, \textit{Open attachment}, \textit{Reply}, or \textit{Unknown}. Five LLMs (\textit{Claude~3.5~Sonnet}, \textit{GPT-5.2-Chat}, \textit{Gemini-2.5-Flash}, \textit{Qwen~2.5-7B-Instruct}, \textit{DeepSeek-Chat}) each labeled all 99 emails five times ($k=5$, temperature~0). Accuracy and internal consistency were measured using a confidence metric adapted from~\cite{10825051}.

\subsection{Step 2.2: Label Augmentation Benchmarking}
\label{sec:label_augmentation}

To populate the missing URL and attachment fields in the aggregated subset, we benchmarked three LLMs (\textit{Claude~3.5~Sonnet}, \textit{GPT-5.2-Chat}, \textit{Gemini-2.5-Flash}) on the same 99 benchmark emails.
Models were instructed to populate missing fields under motivation-aligned structural rules: emails labeled ``Follow the link'' required a non-null URL; ``Open attachment'' required a filename. Category-specific constraints governed domain plausibility. 

For phishing variants, only deceptive look-alike or invented domains were permitted, assuming that Sender Policy Framework (SPF) checks would otherwise flag the use of official domains. Conversely, for spam and legitimate variants, real official domains were allowed, accurately reflecting real-world marketing and promotional practices.
The full prompting strategy and constraint set are detailed in our repository. Outputs were evaluated on quantitative structural correctness and contextual plausibility via manual review. \textit{Gemini-2.5-Flash} was selected based on the highest structural reliability and lowest hallucination rate.

\subsection{Step 3: Label Enrichment}
\label{sec:label_enrichment}

Using the validated prompting strategy, \textit{Gemini-2.5-Flash} populated missing intent (motivation), URL, and attachment fields across all 3,000 aggregated emails. The enriched set was merged with the 300 manually curated emails, yielding a seed dataset of 3,300 structurally consistent emails.

\subsection{Step 4: Dataset Expansion with Seeding}
\label{sec:dataset_expansion}

We expand the dataset using structured LLM-based generation, where each of the $N=3,300$ seed emails serves as an \textit{email template}. For each template $t_i$ ($i = 1, 2, \dots, N$), we have a total of $K=7$ emails (the original seed plus six synthetic variants). We denote the ground-truth label for template $t_i$ as $y_i$, which is shared across all its $K$ instances. 

The six variants (2x3) are generated along two orthogonal dimensions, as shown in Phase 4 of Figure~\ref{fig:methodology}:
\begin{itemize}
    \item \textbf{2 Entity Types:} Globally recognized entities versus fabricated but realistic ones.
    \item \textbf{3 Length Types:} Short (4--8 sentences), medium (10--16 sentences), and long (25--40 sentences). 
\end{itemize}

Six distinct prompt templates cover each entity-length combination, strictly preserving the original email's intent and structural constraints. Consistent with Step~2.2, phishing variants require deceptive domains for well-known entities, while spam and legitimate emails allow realistic corporate domains. This generation process inherently sanitizes encoding artifacts and residual HTML from the aggregated seeds, yielding clean plaintext variants. Additionally, non-English variants were retained in their original languages. This process produced 19,800 synthetic variants, yielding a final evaluation dataset of 23,100 emails.

\subsection{Evaluation and Metrics}
\label{sec:evaluation_metrics}

Qwen-2.5-72B and Gemini-3.1-Pro are evaluated using a single inference pass per email. To measure classification reliability across linguistic variations, we group the predictions by their originating template. Let $q_{i,j}$ denote the predicted label for the $j$-th instance ($j = 1, \dots, K$) of template $t_i$. We adapt the metrics introduced by Tihanyi et al.~\cite{10825051}: 

The \textbf{Task Success Rate ($\mathrm{TSR}$)} counts the number of correctly classified variants for a given template $t_i$ such that:
\begin{equation}
\mathrm{TSR}(t_i, K) = \sum_{j=1}^{K} I\big(q_{i,j} = y_i\big),
\end{equation}
where $I(\cdot)$ is the indicator function returning $1$ if its argument is true and $0$ otherwise, such that $0 \leq \mathrm{TSR}(t_i,K) \leq K$. 

The \textbf{Confidence Index ($\mathrm{Conf}@K$)} measures the percentage of templates that are perfectly classified across all $K$ variants:
\begin{equation}
\mathrm{Conf}@K = \frac{100}{N} \sum_{i=1}^{N} I\big(\mathrm{TSR}(t_i, K) = K\big).
\end{equation}

We additionally introduce the \textbf{Total Flip Score ($\mathrm{TFS}@K$)}, which counts the absolute number of templates where the model consistently fails across all $K$ variants:
\begin{equation}
\mathrm{TFS}@K = \sum_{i=1}^{N} I\big(\mathrm{TSR}(t_i, K) = 0\big).
\end{equation}
By isolating $\mathrm{TSR}(t_i,K) = 0$ cases, $\mathrm{TFS}@K$ pinpoints threat vectors that are reliably evasive, distinguishing fundamental model blind spots from mere stochastic errors.

The two LLMs are evaluated across four dimensions:
\begin{enumerate}
    \item \textbf{Label Configuration:} Three-class classification (Phishing vs.\ Spam vs.\ Valid).
    \item \textbf{Prompting Strategy:} \textit{Basic} (subject, body) vs.\ \textit{Full} (+sender, URLs, filenames).
    \item \textbf{Dataset Condition:} \textit{Original} seed emails vs.\ LLM-\textit{Generated} variants.
    \item \textbf{Source Type:} \textit{Private} inbox collections vs.\ \textit{Public} datasets.
\end{enumerate}


\section{Results}
\label{sec:results}

\subsection{Intent and Label Augmentation Benchmark}

Before expanding the dataset, we benchmarked LLMs on two tasks using 99 manually validated emails: (1) inferring attacker intent and (2) populating missing structural metadata (URLs and attachments).

\textbf{Intent Labeling:} Each model processed the benchmark $r=5$ times per email. To make a single label, we prioritized motivation hierarchically: requests to ``open attachment'' superseded ``follow link''. We report \textit{Strict Accuracy} (all 5 runs match ground truth), and \textit{Consistency} (giving the same answer regardless of correctness). 

\begin{table}[h]
\centering
\caption{Intent Label Results (99 emails, $r=5$ runs).}
\label{tab:motivation_results}
\begin{tabular}{lcc}
\toprule
\textbf{Model} & \textbf{Strict Accuracy} & \textbf{Consistency} \\
\midrule
Claude 3.5 Sonnet & 0.9091 & 0.9838 \\
GPT-5.2-chat      & 0.8788 & 0.9798 \\
\textbf{Gemini 2.5 Flash}  & \textbf{0.9798} & \textbf{1.0000} \\
Qwen 2.5-7B       & 0.6162 & 0.9434 \\
DeepSeek Chat     & 0.8586 & 0.9939 \\
\bottomrule
\end{tabular}
\end{table}

As shown in Table~\ref{tab:motivation_results}, Gemini-2.5-Flash achieved the strongest performance, reaching 97.98\% strict accuracy with perfect internal consistency. 

\textbf{Structural Label Augmentation.} To populate missing fields in legacy emails, we evaluated Gemini 2.5 Flash, GPT-5.2-chat, and Claude 3.5 Sonnet. Quantitative checks verified logical consistency: emails labeled ``follow link'' required a generated URL, and ``open attachment'' required a non-empty filename. Manual review by two experts assessed the plausibility of the generated artifacts. Claude and GPT frequently hallucinated placeholders, generating text like \texttt{[insert link here]} instead of a genuine URL based on the email's narrative. Gemini-2.5-Flash consistently generated realistic, context-aware domains and filenames. 


\subsection{Provenance: Private, Public and their rephrased variants}
\begin{figure}[ht]
    \centering
     \includegraphics[width=1\linewidth]{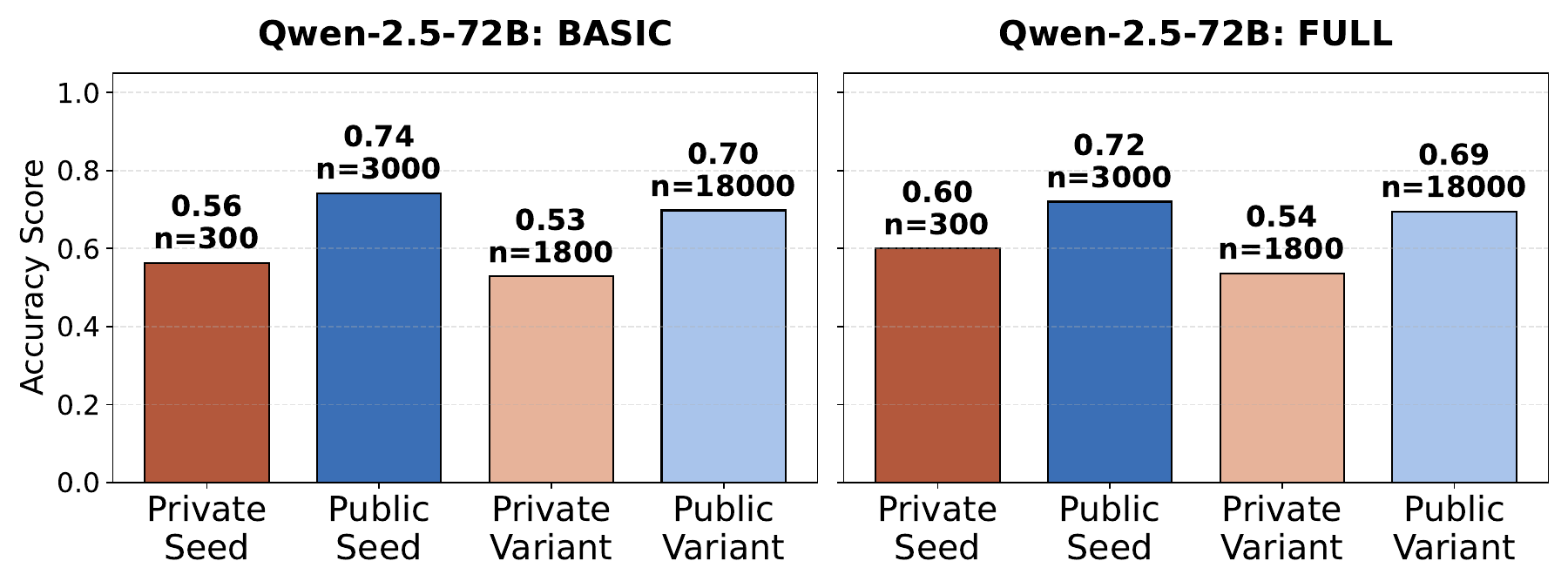}
    \includegraphics[width=1\linewidth]{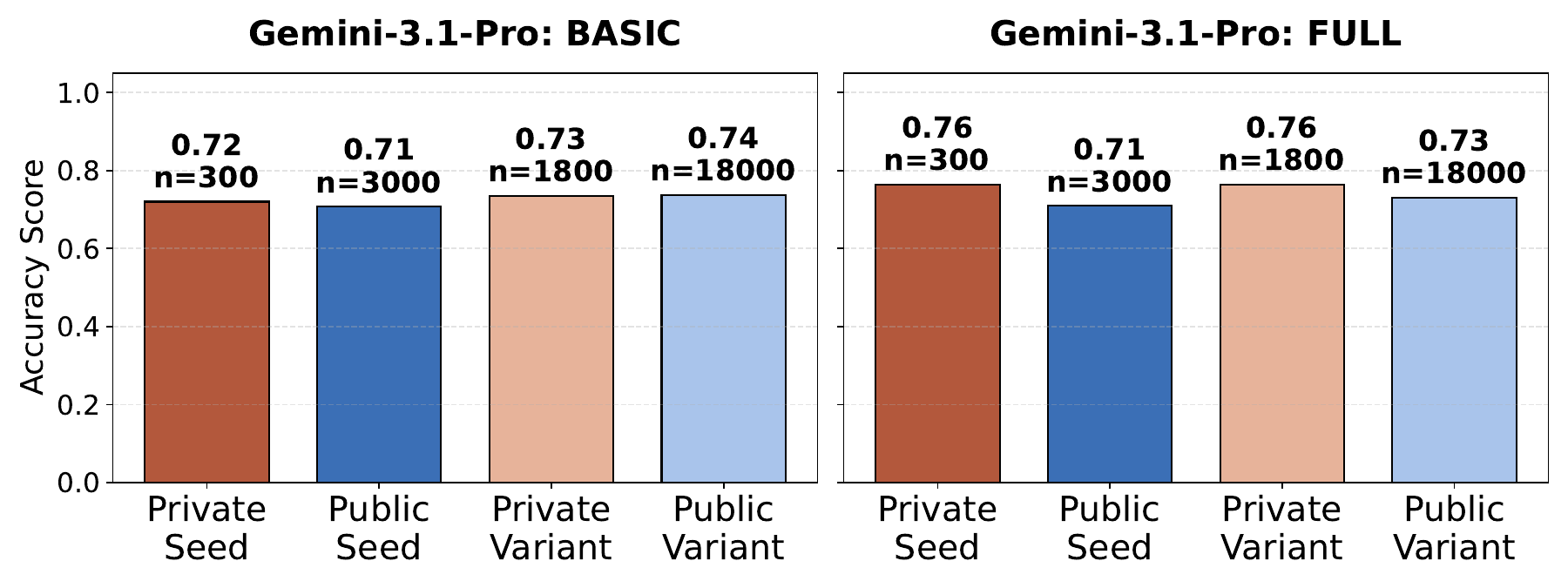}
    
    \caption{Provenance Analysis: Comparison of classification accuracy between BASIC and FULL prompts.}
    \label{fig:prov}
\end{figure}

We track classification accuracy across four distinct data subsets: Private Seeds ($N=300$), Public Seeds ($N=3,000$), Private Variants ($N=1,800$), and Public Variants ($N=18,000$). Figure~\ref{fig:prov} presents these comparisons for Qwen-2.5-72B and Gemini-3.1-Pro, respectively, evaluating both under Basic (body and subject only) and Full (including metadata) prompting configurations. 

\textbf{Private vs.\ Public:} Examining just the Basic setting, Qwen-2.5-72B performs significantly worse on private seed. In contrast, Gemini-3.1-Pro maintains a uniform baseline accuracy across both private and public corpora before matedata is added.

\textbf{Metadata on Provenance:} At first glance, the introduction of metadata benefits private emails more then the public ones.
However, overall accuracy scores can mask underlying class-level shifts, where the recall or F1-score of one class might improve while another drops.  \textbf{When we investigate the confusion matrices, new insights will be gained}. Until then, we cannot draw definitive conclusions from these aggregated figures, to conclude whether the semi-synthetic metadata for Public seed, or the fully synthetic for Private variants and Public variants improve or degrade classification decision.

\textbf{Variants:} On the other hand rephrasing the emails (the variants) had a noticeably different impact on Qwen-2.5-72B and Gemini-3.1-Pro. For Qwen performance drops on variants, suggesting that the ``fuzzing'' is successful and using different wording confuses the classifier, regardless of whether metadata is present. Gemini, on the other hand, shows the opposite: performance on variants improves by 0--2~pp., suggesting that once it understands the email's logic, it stays more consistent with its classification.

\begin{figure*}[t]
    \centering
    \begin{subfigure}[t]{0.48\linewidth}
        \centering
        \includegraphics[width=\linewidth]{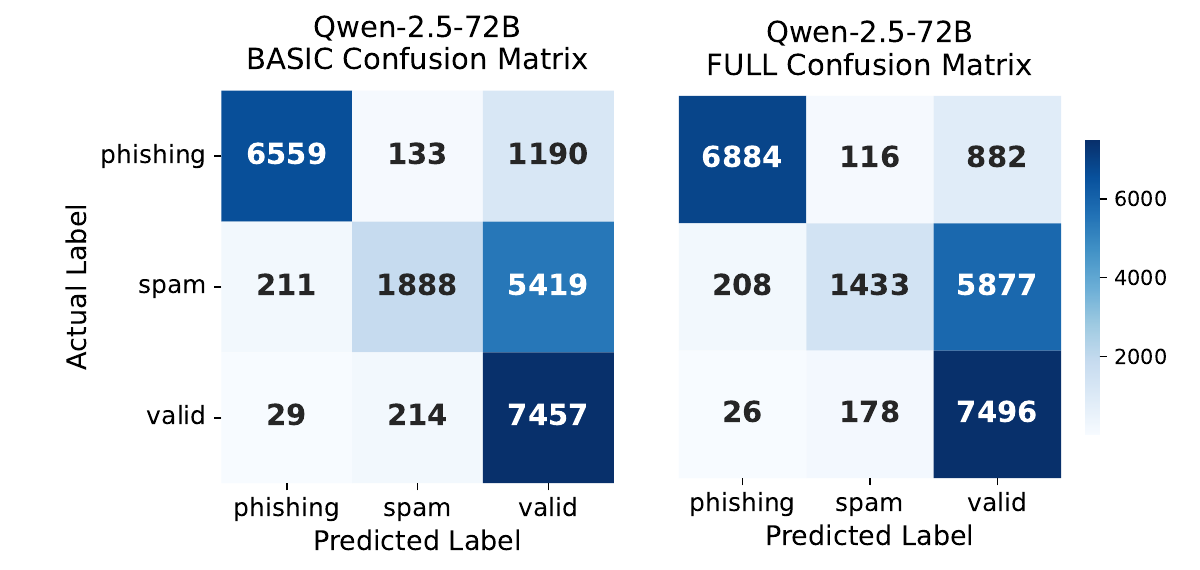}
        \caption{ }
        \label{fig:confusion_qwen}
    \end{subfigure}
    \hfill
    \begin{subfigure}[t]{0.48\linewidth}
        \centering
        \includegraphics[width=\linewidth]{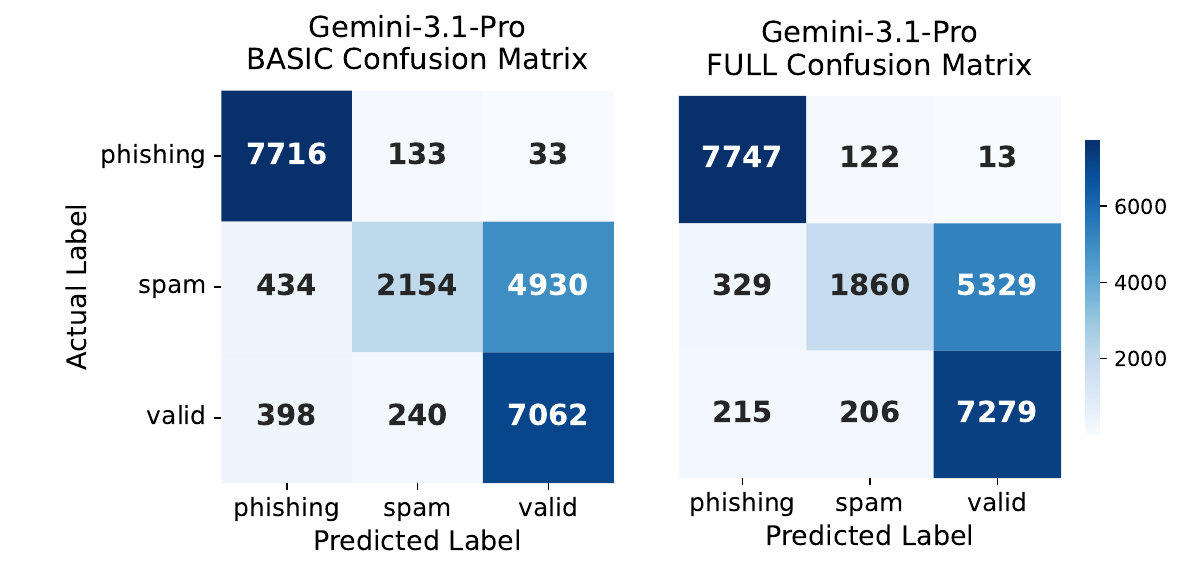}
        \caption{ }
        \label{fig:confusion_gemini}
    \end{subfigure}
    
    \begin{subfigure}[t]{0.48\linewidth}
        \centering
        \includegraphics[width=\linewidth]{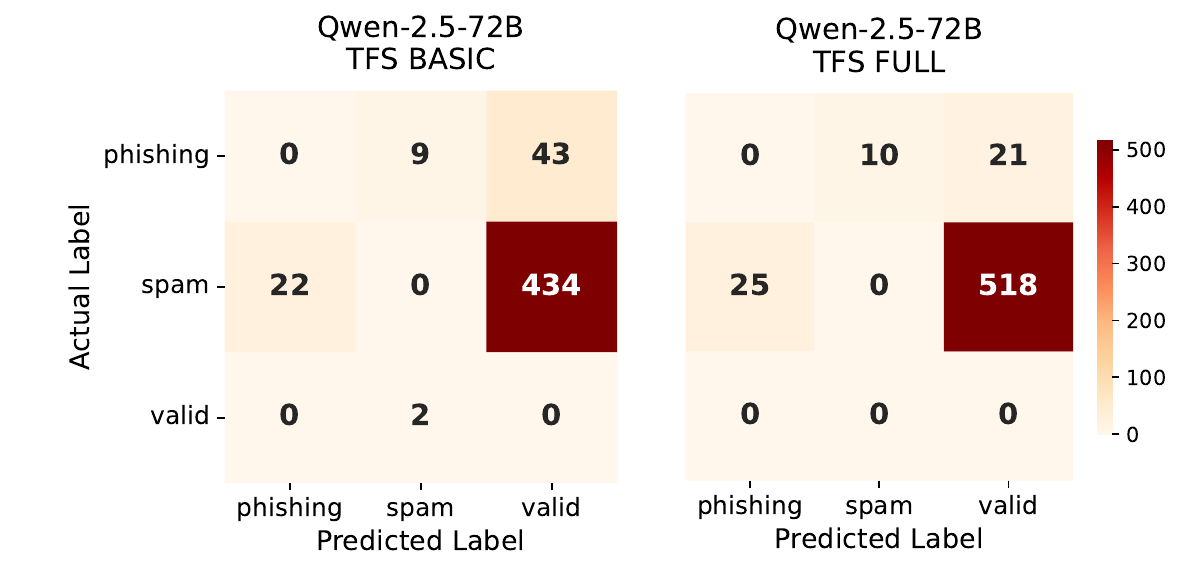}
        \caption{ }
        \label{fig:tfs_qwen}
    \end{subfigure}
    \hfill
    \begin{subfigure}[t]{0.48\linewidth}
        \centering
        \includegraphics[width=\linewidth]{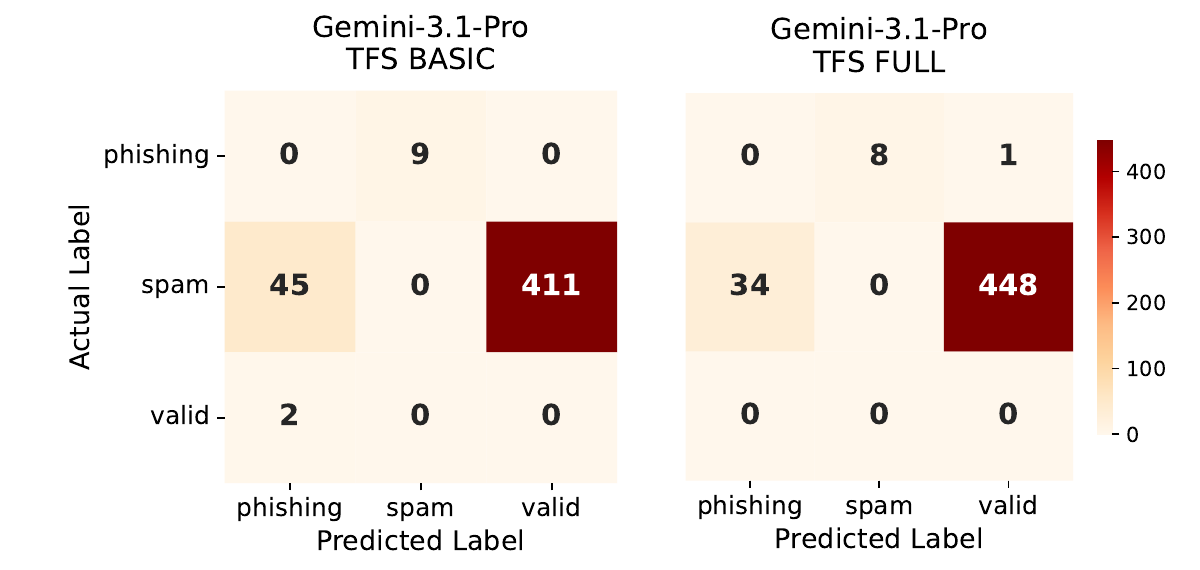}
       \caption{ }
        \label{fig:tfs_gemini}
    \end{subfigure}
    \caption{Confusion matrices (a and b), Total Flip Score Matrices (c and d) under the Basic and Full prompting strategies.}
    \label{fig:combined_all}
\end{figure*}

\begin{table*}[b]
\centering
\caption{Classification Performance and Template Reliability Metrics across Models and Prompts.}
\label{tab:comprehensive_metrics}
\resizebox{\textwidth}{!}{%
\begin{tabular}{ll cccc|cccc}
\toprule
& & \multicolumn{4}{c|}{\textbf{Classification Metrics ($N=23,100$)}} & \multicolumn{4}{c}{\textbf{Template Reliability ($N=3,300$ Templates)}} \\
\textbf{Model} & \textbf{Prompt} & \textbf{Accuracy} & \textbf{Macro F1} & \textbf{F1 (Phish)} & \textbf{F1 (Spam)} & \textbf{\#Conf@7} & \textbf{Conf@7 (\%)} & \textbf{\#TFS@7} & \textbf{TFS@7 (\%)} \\
\midrule
Qwen-2.5-72B   & BASIC           & 0.688 & 0.655 & 0.893 & 0.387 & 1,745 & 52.88\% & 510 & 15.45\% \\
Qwen-2.5-72B   & FULL            & 0.684 & 0.636 & 0.917 & 0.310 & 1,826 & 55.33\% & 575 & 17.42\% \\
\midrule
Gemini-3.1-Pro & BASIC           & 0.733 & \textbf{0.694} & 0.939 & \textbf{0.428} & 1,998 & 60.55\% & \textbf{467} & \textbf{14.15\%} \\
Gemini-3.1-Pro & FULL            & \textbf{0.731} & 0.685 & \textbf{0.958} & 0.383 & \textbf{2,075} & \textbf{62.88\%} & 491 & 14.88\% \\
\bottomrule
\end{tabular}
}
\end{table*}

\subsection{3-Class Performance}

While aggregate accuracy suggests stable performance (Figure~\ref{fig:prov}), the confusion matrices (Figures~\ref{fig:confusion_qwen} and \ref{fig:confusion_gemini}) and corresponding F1 scores (Table~\ref{tab:comprehensive_metrics}) reveal severe class-specific imbalances and a distinct trade-off when introducing metadata.

\textbf{Phishing Detection:} Qwen-2.5-72B acts as a ``relaxed'' classifier, initially misclassifying 1,190 phishing emails as Valid (BASIC); adding metadata reduces this to 882, raising its Phishing F1 from 0.893 to 0.917. Gemini-3.1-Pro demonstrates a stricter posture, misclassifying only 33 phishing emails as Valid, dropping to 13 with metadata, achieving an exceptional Phishing F1 of 0.958.

\textbf{Spam Detection:} Both models struggle to isolate Spam, as reflected by the low Spam F1 scores ($\leq$0.428). Under the BASIC prompt, Qwen-2.5-72B misclassifies 5,419 Spam emails as Valid. Gemini-3.1-Pro similarly marks 4,930 Spam emails as Valid while confusing 434 for Phishing. Counterintuitively, introducing metadata increases these errors. It severely impacts the models' sensitivity to spam, driving Qwen-2.5-72B's already low Spam Recall from 25.11\% down to just 19.06\% and Gemini-3.1-Pro's from 28.65\% to 24.74\%, with the vast majority of these false negatives being absorbed by the Valid class.

\textbf{Valid Classification:} Qwen-2.5-72B's ``relaxed'' nature benefits legitimate emails, incorrectly flagging only 29 Valid as Phishing. Conversely, Gemini-3.1-Pro's ``paranoid'' posture penalizes legitimate traffic, misclassifying 398 Valid emails as Phishing in the BASIC setting. However, metadata provides a crucial corrective signal, reducing Gemini's false positives to 215.

\textbf{The Metadata Trade-off:} The integration of structural metadata forces a zero-sum shift in model behavior. While it is highly beneficial for isolating true Phishing threats and correcting Valid false positives, it systematically pushes borderline Spam into the Valid category. This specific failure mode causes the overall Macro F1 to decline for both models (e.g., Qwen-2.5-72B drops from 0.655 to 0.636) despite metadata improving strict phishing detection.

\subsection{Reliability and Systematic Failure Modes}

\subsubsection{Absolute Blind Spots -- Analysis of Systematic Failures}
To investigate the nature of model weaknesses, we analyze the error distributions of ``Blind Spots'': templates where a model failed to correctly classify the original seed and all six variants ($TFS@7=7$). The resulting Systematic TFS matrices (Figures~\ref{fig:tfs_qwen} and \ref{fig:tfs_gemini}) reveal distinct security postures across these consistently misclassified families.

\textbf{Actual Label - Phishing:} Qwen-2.5-72B misclassified 43 Phishing templates as Valid, though this failure more than halves when metadata is added. Gemini-3.1-Pro, under the Basic setting, exhibits consistent blind spots only when pushing Phishing into Spam. However, adding metadata pushed one Phishing template from Spam into Valid. Upon manual OSINT investigation, this email (\texttt{No. 388}, sourced from the Nazario Dataset) is a legitimate inquiry from an MIT Lincoln Laboratory researcher. We retain the original label for this experiment, but we have corrected it for the published dataset for all variants.

\textbf{Actual Label - Spam:} Both models systematically misclassify Spam templates as Valid, as there is an inherently thin boundary between the two classes. In practice, modern inboxes divide legitimate email into subcategories (Promotions, Updates, Forums) and reserve Spam for aggressive marketing or gray-area communication that fails to fit elsewhere. The models' struggles reflect this real-world ambiguity.

\textbf{Actual Label - Valid:} Systematic misclassification of Valid emails is exceptionally rare. In the Basic setting, Qwen-2.5-72B completely misclassified two templates as Spam, while Gemini-3.1-Pro misclassified two as Phishing. Crucially, introducing metadata eliminated these systematic blind spots for both models, leaving the Valid row empty.

\subsubsection{Model Confidence} 
Table~\ref{tab:comprehensive_metrics} demonstrates that Gemini-3.1-Pro has stronger semantic understanding, allowing it to better interpret a template's logic and remain consistent across variants. Overall, Full prompts containing metadata result in higher model consistency for both classifiers. However, while Qwen-2.5-72B is generally less confident in its correct predictions, it is paradoxically more \textit{confidently wrong} than Gemini-3.1-Pro, systematically failing on 575 templates compared to Gemini's 491 under the Full setting.

\section{Limitations and Future Research}
\label{sec:limit}
This study only benchmarked two LLMs in a zero-shot setting. Future work should compare these results with fine-tuned encoder models such as BERT, and with traditional machine learning classifiers, to test their efficacy in three-class attribution. 
Future work must address the persistent issue of human mislabeling in public benchmark datasets---uncovered during our systematic failure analysis---and explore hybrid architectures that combine LLM semantic reasoning with heuristic filters to better delineate spam from legitimate communication. However, the classification of spam can be highly subjective, as the relevance of promotional emails or conference invitations depends entirely on the recipient's individual preferences. 

\section{Conclusion}
\label{sec:conclusion}

This study evaluates the reliability and systematic failure modes of LLMs in email security classification, focusing on model reliability and the influence of contextual metadata. We distill our findings by formalizing and answering two primary research questions regarding model resilience and feature representation.

\textbf{RQ1: How do state-of-the-art LLMs generalize across unseen private data and rephrased email variants?} \\
Our evaluation reveals a stark contrast between the examined models. Qwen-2.5-72B exhibits strong symptoms of data contamination, demonstrating a notable accuracy degradation on  unseen 2026 emails (both real and synthetic) compared to emails from public datasets. Conversely, Gemini-3.1-Pro maintains robust performance on unseen data. It also shows higher resilience to semantic perturbation ($62.88\%$ confidence index vs. Qwen-2.5-72B's $55.33\%$), suggesting a reliance on underlying logic rather than on brittle lexical patterns. Despite this architectural divergence, both models share a persistent, systematic weakness in differentiating spam from valid emails.

\textbf{RQ2: What is the impact of incorporating structural email metadata on LLM classification performance?} \\
While both models establish strong baselines when relying only on email body and subject, the integration of structural metadata (URLs, sender domains, attachments) decisively enhances phishing detection across architectures---elevating Gemini-3.1-Pro's Phishing F1 to $0.958$ from $0.939$ and Qwen-2.5-72B's from $0.893$ to $0.917$, representing an approximate $20-30\%$ reduction in classification errors. However, this extended context introduces severe class-specific trade-offs by degrading spam detection, causing both models to more frequently misclassify spam emails as valid. Consequently, while metadata is a necessary catalyst for hardening critical phishing defenses, its depressive effect on the overall Macro F1 score underscores the fundamental semantic challenge LLMs face when navigating the subjective boundary between spam and valid communication.

\bibliographystyle{IEEEtran}
\bibliography{references}
\end{document}